\documentclass[aps,prl,twocolumn,nofootinbib,superscriptaddress,10pt]{revtex4-2}
\usepackage[utf8]{inputenc} 
\usepackage{amsmath}
\usepackage{amssymb}
\usepackage[dvipsnames]{xcolor}
\usepackage{graphicx}
\usepackage{bm}
\usepackage{microtype}
\usepackage{mathtools}
\usepackage[colorlinks=true,backref=false, linktocpage=true,
citecolor=blue,urlcolor=blue,linkcolor=blue,pdfpagemode=UseOutlines,pdfstartview=FitH,bookmarksopen]{hyperref}
\usepackage{physics}
\usepackage{enumitem}
\usepackage{balance}

\newcommand{\bes}{\begin{subequations}}
\newcommand{\ees}{\end{subequations}}
\newcommand\sect[1]{\noindent \textbf{\emph{#1.}--}}
\newcommand\appsect[1]{
  \refstepcounter{section}
  \setcounter{equation}{0}
  \noindent \textbf{\emph{#1.}--}
}

\def\a{\alpha}
\def\d{\delta}
\def\g{\gamma}
\def\o{\omega}
\def\l{\lambda}
\def\D{\Delta}
\def\O{\Omega}
\def\tchi{{\tilde{\chi}}}
\def\sA{{\mathcal A}}
\def\sL{{\mathcal L}}
\def\vA{\bm{A}}
\def\vB{\bm{B}}
\def\vE{\bm{E}}
\def\vJ{\bm{J}}
\def\vp{\bm{p}}
\def\vq{\bm{q}}
\def\vv{\bm{v}}
\def\vx{\bm{x}}
\def\pd{\partial}
\def\vpd{\bm\partial}
\def\le{\left}
\def\ri{\right}
\def\no{\nonumber}

\begin{document}

\title{Building Far-from-Equilibrium Effective Field Theories Using Shift Symmetries}

\author{Xin An}
\email{xin.an@ugent.be}
\affiliation{Department of Physics and Astronomy, Ghent University, 9000 Ghent, Belgium}

\author{Robbe Brants}
\email{robbe.brants@ugent.be}
\affiliation{Department of Physics and Astronomy, Ghent University, 9000 Ghent, Belgium}

\author{Michal P. Heller}
\email{michal.p.heller@ugent.be}
\affiliation{Department of Physics and Astronomy, Ghent University, 9000 Ghent, Belgium}

\affiliation{Institute of Theoretical Physics and Mark Kac Center for Complex
Systems Research, Jagiellonian University, 30-348 Cracow, Poland}

\author{Yi Yin}
\email{yiyin@cuhk.edu.cn}
\affiliation{School of Science and Engineering, The Chinese University of Hong Kong (Shenzhen), Longgang, Shenzhen, Guangdong, 518172, China}
\affiliation{Southern Center for Nuclear-Science Theory (SCNT), Institute of Modern Physics, Chinese Academy of Sciences,Huizhou,Guangdong,516000,China},

\begin{abstract}
Contemporary understanding of thermalization in quantum field theory stems largely from understanding properties of transient excitations of equilibria. These nonhydrodynamic excitations are known to structurally differ between weakly and strongly coupled quantum field theories with no known results at intermediate values of the interaction strength. We demonstrate that all the known behaviors of transient excitations can be understood as a consequence of different realizations of a symmetry principle, the shift symmetry, applied at the level of the far-from-equilibrium generalization of the hydrodynamic effective action that we explicitly construct. Our approach naturally includes the effects of stochastic fluctuations outside the hydrodynamic regime and allows to explicitly construct hybrid models interpolating between weak- and strong-coupling behavior. We study properties of one such model motivated by thermalization in nuclear collisions in light of the QCD running coupling.
\end{abstract}

\maketitle

\sect{Introduction} 
Hydrodynamics is often referred to as the universal low-energy effective description of many-body systems near equilibrium~\cite{Landau:2013fluid,Jeon:2015dfa}. Its applications are plentiful and in the context of high-energy physics primarily concern modeling the quark-gluon plasma (QGP) in ultrarelativistic
nuclear collisions at the Relativistic Heavy Ion Collider and
the Large Hadron Collider~\cite{Florkowski:2017olj,Romatschke:2017ejr}. 

Motivated by this experimental research program, studies of the emergence of hydrodynamic behavior in quantum field theories led to a simple yet deep realization: hydrodynamics becomes applicable when excitations not captured by it have decayed~\cite{Berges:2020fwq}. This triggered significant interest in understanding nonhydrodynamic degrees of freedom of holographic quantum field theories~\cite{Horowitz:1999jd,Birmingham:2001pj,Kovtun:2005ev,Buchel:2015saa,Janik:2015waa,Janik:2016btb,Withers:2018srf,Grozdanov:2019kge}, followed by weak-coupling studies in kinetic theory~\cite{Romatschke:2015gic,Kurkela:2017xis,Moore:2018mma,Ochsenfeld:2023wxz,Bajec:2024jez,Brants:2024wrx}. 

Hydrodynamic excitations between the two classes of systems differ only in the associated values of transport coefficients. For example, for the diffusive mode
\begin{align}
\label{diffusion}
    \o= - i D q^2 + {\cal O}(q^4)\,,
\end{align}
the leading difference in hydrodynamic behavior between holography and kinetic theory is encoded in the diffusive constant $D$. The mode frequency $\omega$ receives corrections from higher powers of spatial momentum $q$ that will be also theory-dependent, but the structure itself is not.

In contrast to Eq.~\eqref{diffusion}, nonhydrodynamic sectors are structurally different and diverse. In particular, in strongly coupled quantum field theories, the excitations can be thought of as single-pole singularities of Fourier-transformed retarded correlators of conserved currents and interpreted as transient quasinormal modes (QNMs) of holographic black holes~\cite{Bhattacharyya:2007vjd}. In practical terms, they give rise to an exponential decay in time. In contrast, nonhydrodynamic degrees of freedom in kinetic theory appear as branch cuts or their generalization: nonanalytic regions~\cite{Romatschke:2015gic,Kurkela:2017xis}. The associated decay in time of such nonhydrodynamic sectors can be subexponential (also seen as the long-time tails in fluctuating hydrodynamics~\cite{Kovtun:2003vj,An:2019csj}).

There is now a systematic effective field theory (EFT) for hydrodynamics based on the Schwinger-Keldysh (SK) formalism~\cite{Crossley:2015evo,Crossley:2017,Liu:2018kfw}. This triggers an important open question: can the purview of EFTs be extended to systematically capture the rich landscape of nonhydrodynamic excitations, such as the QNMs in holographic systems or the short-lived excitations in QGP and ultracold gases?

In this Letter we answer this question in a constructive manner utilizing structures inspired on one hand by holography and on the other by kinetic theory. Our study brings four main new results:
\begin{itemize}[left=0pt]
    \item We demonstrate that the shift symmetry known from the earlier studies of hydrodynamic EFTs can be used as a new principle in organizing the description of nonhydrodynamic excitations for a class of theories, including holographiclike and kinetic-theory like theories.
    \item We show that different ways of realizing this symmetry underlies the qualitative difference in excitations in holography vs kinetic theory.
    \item Our approach allows for a construction of a new class of models of nonequilibrium dynamics phrased directly in the language of EFT that are hybrids of holography and kinetic theory. They are motivated by the QCD running coupling and earlier attempts to incorporate it in modeling QGP dynamics~\cite{Casalderrey-Solana:2014bpa,Iancu:2014ava,Mukhopadhyay:2015smb,Kurkela:2018dku,Mondkar:2021qsf}.
    \item The shift symmetry also allows us to systematically describe statistical fluctuations beyond the low-frequency hydrodynamic regime, which so far have eluded attention and understanding.
\end{itemize}
We refer to our framework as the shift-symmetry-based construction. These results open a new pathway for building effective theories for nonequilibrium quantum systems beyond hydrodynamics.

\vspace{10pt}
\sect{Setup}
Our goal is to develop the SK framework for describing both hydrodynamic and relevant nonhydrodynamic excitations and fluctuations. For demonstration, we focus on the sector with a conserved U(1)$_V$ charge current $J^{\mu}$ in a symmetry-unbroken phase. 

The construction of SK action involves three key steps~\cite{Crossley:2015evo,Liu:2018kfw,Akyuz:2023lsm}. 
The first step is to choose suitable dynamical fields, collectively denoted by $\chi^{r,a}$, in the $r$-$a$ basis~\cite{Keldysh:1964ud} where $r$-fields represent physical observables and $a$-fields encode fluctuations.
Second, one must identify the symmetry of the action, both the microscopic symmetry, such as U(1)$_V$, and importantly, the emergent symmetries that arise from the state under consideration.
Finally, the action must be written down in a way that is consistent with these symmetries and other physical constraints (cf. Eq.~\eqref{I-unitary} below), organized via a gradient expansion scheme. 

In the EFT for diffusion~\cite{Crossley:2015evo,Haehl:2015foa,Jensen:2017kzi,Crossley:2017,Chen-Lin:2018kfl},  
the U(1)$_V$ phase fields $\chi=\psi,\chi^{a}=\psi^{a}$ (omitting the $r$ subscript from now on) serve as dynamical variables for describing the evolution of the associated conserved charge density.
This choice is particularly convenient for implementing U(1)$_V$ symmetry. Additionally, the action must be invariant under 
spatial shift transformation of the ($r$-)phase field~\cite{Crossley:2015evo}
\begin{align}
\label{shift-H}
\d_{s}\psi=\lambda(\vx)\,,
\end{align}
while $\d_{s}\psi^a=0$. This emergent \emph{shift symmetry} is proposed to ensure that the fluid under study is in the symmetry-unbroken state, as it allows independent phase choices at different locations $\vx$~\cite{Liu:2018kfw,Akyuz:2023lsm}.
It prohibits unwanted propagating Goldstone modes (eliminating terms linear in $q$ in Eq.~\eqref{diffusion}) in the EFT thus constructed, which correctly generates diffusive dynamics and fluctuations~\cite{Chen-Lin:2018kfl}.

The central ingredient of our construction is the generalization of the shift symmetry underlying the EFT for diffusion. 
We propose that multiple nonhydrodynamic excitations can be described by the action of a system with multiple global U(1) symmetries. 
Some of those symmetries can be spontaneously broken and subsequently gauged to local ones. 
The dynamical fields include the phase fields of U(1)s and the dynamical gauge fields when the broken symmetries are gauged.
There are three generic scenarios:
\begin{itemize}[left=5pt]
    \item [a)] all U(1) symmetries are unbroken;
    \item [b)] all U(1) symmetries are spontaneously broken into a diagonal U(1);
    \item[c)] the mixtures of a) and b).
\end{itemize}
For each case, 
we formulate the appropriate and nontrivial shift symmetries (cf. Eqs.~\eqref{shift-UB} and \eqref{shift-B}) and demonstrate that the difference in those shift symmetries results in qualitatively distinctive features in excitations.

The action additionally satisfies several physical conditions~\cite{Crossley:2017,Liu:2018kfw}
\small
\begin{equation}
\label{I-unitary}
I[\chi,\chi_{a}=0] = 0\,,\quad
I[\chi,-\chi_{a}] =-I^{*}[\chi,\chi_{a}]\,,\quad
\mathrm{Im}\, I \geqslant 0\,.
\end{equation}
\normalsize
The first condition in Eq.~\eqref{I-unitary} implies that the action can be organized as a series expanded in $a$-field: $I=I_a+I_{aa}+\ldots$ where the subscript counts the power in $\chi_a$. 
We shall first construct $I_{a}$, which suffices to determine excitations. 
Then we return to $I_{aa},$ which encodes Gaussian fluctuations, and demonstrate how shift symmetry is crucial for describing these fluctuations.

\vspace{10pt}
\sect{Unbroken U(1) and kinetic theory}
We begin by formulating the shift symmetry for scenario a) where the phase fields associated with multiple global U(1)$_\alpha$ are denoted by $\phi_{\a}$ with $\a=1,\ldots,N$.
To ensure that those U(1)s are unbroken,
we propose that the action in this scenario, $I_{\rm U}$, 
is invariant under $\alpha$-dependent spatial shift transformation
\begin{align}
   \label{shift-UB}
       \d_{s}\phi_{\a}= \l_{\a}(\vx)\, ,
\end{align}
which naturally generalizes the transformation~\eqref{shift-H}. 
We further require the action to be invariant under U(1)$_V$ transformations $\phi_\alpha \to \phi_\alpha + \Lambda_V(x)$, allowing for a vector current $J^\mu$ obtained via variation w.r.t. the background gauge field $V_\mu$.

With this, we can construct the most general $I_{\rm U}$.
For now, we highlight a specific but physically important case where $N\to \infty$ and the discrete index $\a$ is replaced by the single-particle velocity~$\vv$ (or momentum $\vp$). 
In this limit, $I_{\rm U}$ recovers the action that describes collisionless linearized kinetic theory~\cite{Huang:2024uap} (see Ref.~\cite{Delacretaz:2022ocm} for the nonlinear cases) for the fluctuation of single-particle distribution $n(x,\vp)$ around a homogeneous and static background $n_0(\epsilon_0)$, where the single-particle energy $\epsilon_0$ and $\vv=d\epsilon_0/d\vp$ depend on $\vp$.
Explicitly (in the absence of $V_{\mu}$),
the action symmetric under \eqref{shift-UB} reads
\begin{align}
\label{I-Bolt}
    I_{{\rm U},a}=\int_{x,\vp}
    &\,[-n'_0(\epsilon_0)]
    \Big\{
    [\pd_{t}\phi(\vv)]\pd_{t}\phi^{a}(\vv)
    \no\\
&+\frac{1}{2}\left[(\vv\cdot\vpd)\phi(\vv)\,\pd_{t}\phi^a(\vv)+(r\leftrightarrow a)\right]
    \Big\}\,.
\end{align}
The equation of motion (EOM) from Eq.~\eqref{I-Bolt} for $\phi(\vv)$ yields the standard linearized kinetic equation with the identification $\d n\equiv n-n_0(\epsilon_0)=-n'(\epsilon_0)\,\pd_t\phi(\vv)$, see l.h.s. of Eq.~\eqref{phi-EoM-H1}.
The $\vv$-dependent shift symmetry reflects that particles with different velocities choose their phases independently. 
The gauge-invariant form in the presence of $V_{\mu
}$ can be found in Refs.~\cite{Delacretaz:2022ocm,Huang:2024uap}.

\vspace{10pt}
\sect{Broken Hidden U(1)}
We now examine scenario b) where the symmetry group U(1)$_1\times\ldots$U(1)$_{K}\times$U(1)$_V$ is spontaneously broken into a diagonal U(1). 
The relevant phase fields are $\varphi_{n}$ with $n=1,\ldots,K$ and $\varphi_V$.
Unlike scenario a), we can gauge $\varphi_n$ by introducing dynamical ``photon'' fields $A_{n,\mu}$ and their a-field counterpart which transform with independent gauge parameter functions.
This does not lead to unwanted gapless excitations since those gauge fields acquire mass through the Higgs mechanism.
The gauge-invariant dynamical variables include $\sA_{n,\mu}\equiv (\pd_{\mu}\varphi_{n}+A_{n,\mu})$ and $\sA_{K+1,\mu}\equiv(\pd_{\mu}\varphi_{V}+V_{\mu})$.

For the ``broken'' scenario, we propose the shift symmetry under the following transformation
\begin{align}
\label{shift-B}
   \d_{\rm s}\varphi_{n}=\l(t,\vx)\, \qquad
   \text{so}\, \qquad\d_{s}\sA_{n,\mu}=\pd_{\mu}\l(t,\vx)\,. 
\end{align}
Since U(1)$_n$s are broken, we can no longer require the shift function to depend on index $n$. Compared to Eq.~\eqref{shift-UB}, the shift function $\l(t, \vx)$ varies in time, and the transformation~\eqref{shift-B} also applies to the gauge-invariant combination~$\sA$. 
To satisfy this shift symmetry, the action will depend on the phase fields only through shift-invariant combinations $\sum_{n} c_n\varphi_n$ with $K$ independent choices of $c_n$ such that $\sum_{n} c_n = 0$. We can therefore decompose them into the basis of Goldstone bosons $\D\varphi_n \equiv \varphi_{n+1}-\varphi_n$.

Up to second order in derivatives, the action, expressed in terms of $\D \sA_{n,\mu}\equiv \sA_{n+1,\mu}-\sA_{n,\mu}$ and ``EM'' fields $\vE_{n}={\bm \nabla}A_{n,t}-\pd_{t}{\bm\vA}_n$ and $\vB_n={\bm \nabla}\times \vA_n$, reads
\begin{align}
\label{I-HLS}
    I_{{\rm B},a}[\varphi_V,\sA_n]=&\int_x\sum^{K}_{n=1}\,
     \Big(
     f^2_n\,\D \sA_{n,t}\,\D \sA^{a}_{n,t}- g^2_{n} \D{\bm \sA}_{n}\cdot\D{\bm \sA}^{a}_{n}
    \nonumber\\
    &+\epsilon_n \vE_n\cdot\vE^{a}_n-\kappa_n\vB_{n}\cdot\vB^{a}_n 
     +\sigma_{n}\vE_{n}\cdot {\bm\sA}^a_{n}
     \Big)\, , 
\end{align}
where we count $\D\sA$ the same order as $\partial\sA$ and focus on situations where $\sA_{n}$ can only couple to its neighbors $\sA_{n\pm 1}$. This choice is motivated by the locality of interactions in terms of an additional spatial dimension, along which the $K$ layers lie~\cite{Arkani-Hamed:2001kyx,Son:2003et}. By employing field redefinition, we remove terms of the form $\Delta {\bm\sA}\cdot \vE$.
The above action contains the dissipative part $I_{a,{\rm diss}}$ that is proportional to conductivity $\sigma_{n}$.
The remaining nondissipative part can be factorized as the difference between the action on the forward and backward leg of the SK  contour~\cite{Liu:2018kfw}. 
This part depends on the ``decay constant'' $f_n$, the gauge boson mass $g_n$, and the
electric and magnetic constants, $\epsilon_n, \kappa_n$ for each hidden photon sector.

We observe that if a finite background charge density is permissible, we should not treat $\sA_{1,t}$ as a dynamical field, but fix it to be $\sA_{1,t}=0$ (see End Matter).
Under this condition, the solutions to the EoM only obey the spatial shift, consistent with the invariance of the hydrodynamic action under the spatial shift symmetry~\cite {Liu:2018kfw}: when integrating out those hidden photon fields in the hydrodynamic limit, our theory has to reduce to hydrodynamics.
The EOM for $\varphi_V$ expresses charge conservation
\begin{align}
\label{Eqn-psiV}
\pd_{\mu}J^{\mu}_{\rm B}=0\, ,
\qquad
J^{\mu}_{\rm B}= (f^2_{K}\Delta \sA_{K,t},-g^2_{K}\Delta {\bm \sA_{K}})\,,
\end{align}
where the charge current for the broken hidden symmetry scenario $J^{\mu}_{B}$ is obtained by varying the action w.r.t. $V_{\mu}$. 
The $\sA_{n,\mu}$ obey Maxwell-like equations in the medium: 
\small
\bes
\label{H-Maxwell}
\begin{align}
\label{At-EoM}
\hspace{-6 pt}\epsilon_{n}\vpd\cdot \vE_n &= 
-f^2_{n}\Delta\sA_{n,t}+f^2_{n-1}\Delta\sA_{n-1,t}\,,
\\
\label{Ai-EoM}
\hspace{-6 pt}    \epsilon_{n}\pd_{t}\vE_{n}-\kappa_{n}\vpd\times\vB_n &=
    -g^2_{n}\Delta\sA_{n}+g^2_{n-1}\Delta\sA_{n-1}
    -
\sigma_{n}\bm{E}_{n}.
\end{align}
\ees
\normalsize
For \eqref{At-EoM}, $n=2,\ldots,K$ since $\sA_{1,t}$ is nondynamical while for Eq.~\eqref{Ai-EoM}, $n=1,\ldots, K$.

We focus on excitations in the longitudinal channel  and express Eq.~\eqref{Eqn-psiV}, at $V_{\mu}=0$, as $ R(\o,\vq)(\o\varphi_{V})=0$ where $R(\o,\vq)$ is determined by solving the EOM for $\sA$.
The~excitations are then determined by $R(\omega,\vq)=0$. 
Below, we survey various examples.

\vspace{10pt}
\sect{The $K=1$ and Maxwell-Cattaneo theory}
When $K=1$, the dispersion relation is given by (omitting the subscript $n=1$) 
\begin{align}
\label{eq:Dis-1}
R(\o,\vq)= f^{2}
\left(
\o+\frac{u^2\,q^2(\epsilon\,\o+ig^2\tau_{\rm B})}{g^2-\o(\epsilon\,\o+i g^2\,\tau_{\rm B})}
\right)
=0\,,
\end{align}
where $\tau_{\rm B}\equiv \sigma/g^2, u\equiv g/f$. 
For small $q$,
Eq.~\eqref{eq:Dis-1} yields a diffusive mode~\eqref{diffusion}
with diffusive constant $D_{\rm B}=u^2\tau_{\rm B}$ and a pair of nonhydrodynamic modes $\o_{\pm}=\pm \o_{R}-i \o_{I}$ determined by 
$
\epsilon\o^2+i g^2\,\tau_{\rm B}\,\o-g^2=0
$,
where the magnitude of $\o_{R}$ is set by the gauge boson mass $g$. 
We refer to such excitations with nonzero $\mathrm{Re}(\omega)$ at $q=0$ as \emph{damped oscillatory {\rm (DO)} modes}, which resemble holographic QNMs. 
Setting $g=\sqrt{2\epsilon}/\tau_{\rm B}$ gives $\omega_{\pm}=(\pm 1-i)\tau^{-1}_{\rm B}$.
With $\tau_{\rm B}=1/(2\pi T)$ and $u=1$, we recover the first pair of QNMs $\omega_{\pm}=(\pm 1-i)2\pi T$ and the diffusion constant $D_{\rm SYM}=1/2\pi T$ of strongly coupled $\mathcal{N}=4$ supersymmetric Yang-Mills (SYM) theory~\cite{Kovtun:2005ev,Bu:2015ame}, indicating the model's capability and flexibility to describe holographic excitations. This discussion is reminiscent of linear regime of relativistic hydrodynamics EOMs considered in~\cite{Heller:2014wfa,Heller:2021yjh} with the key difference being that the present discussion is driven by the symmetry principle.

The special case $\epsilon=\kappa=0$ is also interesting.
The Lagrangian density, in the absence of external fields, reduces~to
\begin{multline}
\label{MC}
    \sL_{{\rm B},K=1}=\,f^2
    \Big[(\pd_t\varphi_V)\pd_t\varphi^a_V-u^2\,(\vpd\varphi_V-{\bm \sA})\cdot (\vpd\varphi^a_V-{\bm \sA}^a)
    \\ 
    -u^2\,\tau_{\rm B}\,{\bm\sA}^{a}\cdot\pd_{t}{\bm\sA}\Big]\,.
\end{multline} 
At small $q$, there is only one nonhydrodynamic mode which is purely imaginary, while the diffusive one is unchanged.
At large $q$, the excitations are a pair of DO modes with phase velocity $u$. 
This behavior is characteristic of Maxwell-Cattaneo (MC) theory~\cite{Liu:2024tqe,Ahn:2025odk}.
Indeed,
the EOM for ${\bm \sA}$ from Eq.~\eqref{MC} can be recast into the equation for $\vJ_{\rm B}=-f^2u^2\,(\vpd\varphi_V-{\bm\sA})$, which reads
$\tau_{\rm B}\pd_{t}\vJ_{\rm B}= -\le(\vJ_{\rm B}+D_{\rm B}\vpd J^{t}_{\rm B}\ri)$.
This is precisely the MC equation for the spatial current, which relaxes to Fick’s law at the timescale $\tau_{\rm B}$,  
and is analogous to Müller-Israel-Stewart theory in the energy-momentum sector~\cite{Florkowski:2017olj}.
This demonstrates that Eq.~\eqref{MC} provides an action formulation of MC theory.

\vspace{10pt}
\sect{General $K$ and holography}
For general $K$, there are K pairs of DO modes in nonhydrodynamic sectors, corresponding to the propagation and dissipation of ``massive photons''. 
The continuum limit $K \to \infty$ is of particular interest. Introducing an extradimensional coordinate $\rho$ by $\rho_n= n\,\Delta \rho$ (with $\Delta \rho$ infinitesimal) and viewing $\sA_{n}=\sA(\rho_n)$, 
the action~\eqref{I-HLS} becomes that of Maxwell theory in $4+1$ dimensional curved background (see End Matter). 
Thus, holographic models can be viewed as a special class of action~\eqref{I-HLS} in the continuum limit. 
This relationship is analogous to that between hidden local symmetry models and holographic QCD~\cite{Son:2003et}.

Previous works have coupled holographiclike QNMs to hydrodynamic sectors~\cite{Heller:2014wfa} and used MC-type equations to ensure causality. 
The action~\eqref{MC} was studied by Nickel and Son in the context of holographic liquids~\cite{Nickel:2010pr}. 
However, the unified framework presented here, which describes diverse nonhydrodynamic excitations, is new in the literature.

\vspace{10pt}
\sect{Shift symmetry and structural difference in excitations}
Our framework offers new insights into the differences in the analytic structures of kinetic and holographic theories. This distinction becomes evident when we consider the homogeneous limit \(q \to 0\), where holographic QNMs behave as DO modes, while kinetic excitations are purely imaginary. 
This difference is often attributed to variations in interaction strength. 

However, our analysis reveals that this distinction is closely tied to the underlying symmetry structure. In scenario b), the ``broken case'' where holographic theories belong, the \(n\)-independent shift symmetry allows for a mass term \(\Delta{\bm{\mathcal{A}}}_{n}\cdot\Delta {\bm{\mathcal{A}}}_{n}^a\) in \(I_{\rm B}\). This term generates the real part in the dispersion relation.
Conversely, in scenario a), the "unbroken case", the transformation prohibits the analogous term \(\Delta\phi_{\alpha}\Delta\phi_{\alpha}^{a}\) in \(I_{\rm U}\), thereby resulting in a purely imaginary spectrum at \(q=0\).

\vspace{10pt}
\sect{The hybridization of kinetic theory and QNMs}
Let us consider the mixed scenario.
One motivation is to explore the properties of QGP.
It is anticipated that QGP behaves like a holographic liquid at long distances while admitting a kinetic description with quasiparticles at shorter scales, making hybrid approaches particularly relevant~\cite{Casalderrey-Solana:2014bpa,Iancu:2014ava,Mukhopadhyay:2015smb,Kurkela:2018dku,Mondkar:2021qsf}.

A naive combination of the kinetic action $I_{\rm U}$ with $I_{\rm B}$ conserves charge separately in each sector, yielding two diffusive modes rather than one. 
This doubling of hydrodynamic modes when hybridizing holographic and kinetic sectors has already been observed in previous models~\cite{Kurkela:2018dku}.
Instead, we couple those sectors through $\varphi_{V}$ as
$I= I_{{\rm U}}[\phi]+I_{\rm B}[\varphi_{V},\sA]+ \Delta I[\phi,\varphi_{V}]$ where
\begin{align}
\label{L-diss}
    \Delta\sL= -\frac{f^2_{\rm U}}{\tau_{\rm R}}\int_{\O}
    (\pd_t{\phi}(\vv)-\pd_{t}\varphi_{V})({\phi}^a(\vv)-\varphi^a_{V}) \,.
\end{align}
Here $\int_{\O}=\int d\O/4\pi$ represents the solid angle integration in phase space and $f_{\rm U}^2=\int \frac{dp}{2\pi^2}p^2[-n'(\epsilon_0)]=\chi_{\rm U}$ is the kinetic sector susceptibility. 
The combination preserves gauge invariance and shift symmetry for $\phi(\vv)$. 

From this action the EOM for $\phi(\vv),\varphi_{V}$ read
\bes
\label{phi-EoM-H}
\begin{align}
\label{phi-EoM-H1}
\,&  (\pd_{t}+\vv\cdot\bm\partial) [\pd_{t}\phi(\vv)]
=-\frac{1}{\tau_{\rm R}}(\pd_{t}\phi(\vv)-\pd_{t}\varphi_{V})\,,
     \\
\label{phi-EoM-H2}     
\,&    \pd_{\mu}J^{\mu}_{\rm B}=-
\frac{f_{\rm U}^{2}}{\tau_{\rm R}}\int_{\O}(\pd_{t}\varphi_{V}-\pd_{t}\phi(\vv))\,.
\end{align}      
\ees
It follows that the total charge is conserved $\pd_\mu (J_{\rm B}+J_{\rm U})^\mu=0$ where the charge current for the kinetic sector is given by $J_{\rm U}^{\mu}=f_{\rm U}^2\int_{\Omega}\, (1,\vv)\, \pd_{t}\phi(\vv) $.
The charge density will redistribute between the two sectors until they reach a chemical equilibrium where
$\mu_{\rm B}\equiv\pd_{t}\varphi_{V}= \int_{\O}\pd_{t}\phi(\vv) \equiv \mu_{\rm U}$ such that the r.h.s. of Eq.~\eqref{phi-EoM-H2} vanishes. 
The charge ratio after chemical equilibration is $r\equiv [J_{\rm U}^t/J_{\rm B}^t]_{\rm eq} = \chi_{\rm U}/\chi_{\rm B}$, with $\chi\equiv J^t/\mu$ the susceptibility, and determines the relative importance of the two sectors. 
Because of this chemical equilibration process, 
we find one diffusive mode $\o_{\rm D}$ and one damping mode $\o_{\rm C}$ instead of two diffusive modes. 
In particular, the damping mode is given by $\o_{\rm C}=-i(1+r)/\tau_{\rm R}$, indicating that the chemical equilibration between two sectors is faster than the thermalization of quasiparticles. Note that the susceptibilities and hence 
$r$ are all temperature-dependent.

\begin{figure}[t]
 \centering
  \includegraphics[width=0.4\textwidth]{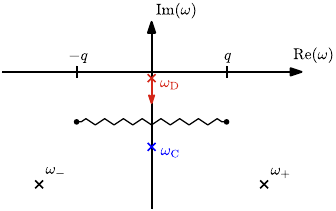}
  \caption{Mode structure of the retarded correlation function for the hybrid model at small values of $q\tau_{\rm B}$, showing the hierarchy between the excitations. The arrow indicates the motion of the diffusive mode when increasing $q$.}
  \label{fig:hybrid-QNM}
\end{figure}

Explicitly solving Eq.~\eqref{phi-EoM-H} along with Eq.~\eqref{H-Maxwell}, the condition for the dispersion relations reads (assuming relativistic particle in the kinetic sector)
\begin{align} 
\label{eq:hybrid dispersion}
    R(\o,\vq)+\frac{i f_{\rm U}^2}{\tau_{\rm R}}\left(1-\frac{1}{2i q\tau_{\rm R}}\log
    \le(
    \frac{\o-q+i\tau^{-1}_{\rm R}}{\o+q+i\tau^{-1}_{\rm R}}
    \ri)\right)=0\, .
\end{align}
Without $R$, it reduces to the condition for excitations in RTA kinetic theory~\cite{Romatschke:2015gic}. 
For definiteness, we now take $R(\o,\vq)$ for $K=1$ from Eq.~\eqref{eq:Dis-1} and illustrate the structure of excitations in Fig.~\ref{fig:hybrid-QNM}.
The branch points $\pm q-i\tau_{\rm R}^{-1}$ originated from the branch cut associated with quasiparticle excitations, while the two DO modes, $\o_{\pm}$, come from the broken sector. 
At small $q$, the diffusive mode is
\begin{align}
\label{D-hybr}
    \omega_{\rm D} =-\frac{i}{1+r}\left(u^2\tau_{\rm B}+\frac{r}{3}\tau_{\rm R}\right)q^2 \,,
\end{align}
so the diffusivity interpolates between $D_{\rm B}=u^2\tau_{\rm B}$ (small~$r$) and kinetic theory's $\tau_{\rm R}/3$ (large $r$).

By matching the broken sector to strongly coupled gauge theories, we find ${\rm Im} \,\omega_{\pm} \sim \tau_{\rm B}^{-1} \sim T$, while the kinetic description of QGP gives $\tau_{\rm R} \sim 1/(g_{\rm QCD}^4 T)$. This naturally establishes the hierarchy $\tau^{-1}_{\rm R} \ll \tau_{\rm B}^{-1}$, implying that quasiparticles dominate at timescales much longer than $\tau_{\rm B}$. 
Nevertheless, for small (but not exponentially small) $r$, the diffusive constant~\eqref{D-hybr} can still be comparable to~$D_{\text{SYM}}$. 
Despite its simplicity, the model presents a novel scenario to resolve how QGP can exhibit both quasiparticle excitations and strongly coupled transport. 

\vspace{10pt}
\sect{Fluctuation and shift symmetry}
The action formalism is advantageous for describing fluctuations, which are captured in terms beyond the linear order of the a-field.
We now present the construction of $I_{aa}$. 
The fluctuation-dissipation (FD) relation [or Kubo–Martin–Schwinger (KMS) condition] fixes $I_{\rm aa}$. 
This can be implemented by requiring invariance of $I_{\rm FD} \equiv I_{a,{\rm diss}} + I_{aa}$ under the KMS transformation $\chi_{\pm}(t) \to \Theta\chi_{\pm}(t \pm i/2T)$, where $\chi_{\pm} = \chi \pm \chi_a/2$ represent forward($-$) and backward ($+$) contour fields.
However, this transformation is nonlocal, indicating $I_{aa}$ thus obtained is also nonlocal. 
The locality of KMS transformation is only restored at low-frequency $\omega/T\ll 1$ where   $\chi(t)\to\Theta\chi(t), \chi^a(t)\to\Theta\chi^a(t)+iT^{-1}\pd_{t}\Theta\chi(t)$, but nonhydrodynamic modes are generally important when $\o/T$ is not small. 

Inspired by the successful description of the tower of QNMs through inclusion of ``hidden'' fields, we introduce a continuum of fields $\tilde{\chi}(\theta, x)$ with an extra coordinate $\theta \in (0, 2\pi)$ for the construction of $I_{aa}$. 
Including $\tchi$ is also motivated by the general situation in field theory, where recovering locality requires relevant modes to be ``integrated-in''.
Those fields \(\tilde{\chi}(\theta, x)\) are assumed to be related to SK fields via
\begin{align}
\label{chi-bound}
 \tchi(\theta=0)=\chi_{-}\,,
 \qquad
 \tchi(\theta=2\pi)=\chi_{+}\, . 
\end{align}
We seek a local action $\tilde{I}[\tilde{\chi}]$ such that
\begin{align}
\label{Path-relation}
    \exp\left(i\,I_{\rm FD}[\chi,\chi_{a}]\right)=\int {\cal D}\tilde{\chi}\, \exp\left(i\tilde{I}[\tilde{\chi}]\right)\,. 
\end{align}

The key idea underlying the construction is again the shift symmetry of $\tilde{I}$ under the transformation:
\begin{align}
\label{shift-M}
    \d_{s}\tchi(\theta,x)=\tilde{\l}(\bm{x})\,.
\end{align}
We take this spatial shift symmetry as our postulate and write down the most general quadratic action that is consistent with Eq.~\eqref{I-unitary} and up to second order in derivatives
\begin{align}
\label{I-flut}
 \tilde{I}= \int_{x}\int^{2\pi}_{0} d\theta\,\gamma \left[2\pi T' i(\pd_{\theta}\tchi)^2- \,(\pd_{\theta}\tchi)\,\pd_{t}\tchi
    \right], 
\end{align}
where $\g,T'$ are real parameters, and the $i$ factor in front of $(\pd_{\theta}\tilde{\chi})^2$ together with $\g\geq 0$ is required by Eq.~\eqref{I-unitary}. The $(\pd_{t}\tilde{\chi})^2$ term is admissible by symmetry but is excluded by the condition $I[\chi,\chi_a=0]=0$ in Eq.~\eqref{I-unitary} since this term does not vanish when $\tilde{\chi}(\theta,x)=\chi(t)$. Evaluating the path integral~\eqref{Path-relation} at the saddle point under the condition~\eqref{chi-bound} gives
\begin{align}
\label{I-FD-full}
I_{\rm FD}=\int_{\vx,\o}\,i\,\o\,\g\,\chi^a(-\o)\left[
\chi(\o)+\frac{1}{2}\coth\left(\frac{\omega}{2T'}\right)\chi^{a}(\o)
\right]\,.
\end{align} 
The action~\eqref{I-FD-full} is invariant under the KMS transformation with $T'=T$.
Note that the first term in real space $\int d^4x(-\g\chi^a\pd_{t}\chi)$ represents damping, but the second one, corresponding to $I_{aa}$, is nonlocal. 
This concludes our construction, where the nonlocal $I_{\rm FD}$ is represented by the local action~\eqref{I-flut} in terms of the hidden fields $\tchi$ obeying shift symmetry and condition~\eqref{chi-bound}.  
To our knowledge, both the result and the shift-symmetry-based approach we employ are new in the literature.
Notably, the derivation of the action~\eqref{I-flut} is solely based on symmetry, making it potentially useful in far-from-equilibrium situations. 

Our general approach to fluctuations applies to various cases. 
For the action~\eqref{I-HLS}, 
identifying ${\bm{\sA}},\sigma$ with $\chi,\g$ respectively yields a complete description of excitations and fluctuations. Recent efforts~\cite{Jain:2023obu,Mullins:2023tjg,Mullins:2025vqa} have incorporated fluctuations in MC-type theories using local KMS symmetry, while fluctuating RTA kinetic theory was similarly formulated in~\cite{Abbasi:2025teu}.
Our approach goes beyond them by remaining valid for general frequencies.
In the continuum limit, our approach matches with the SK action for holographic theories, which are based on a specific prescription to select the in-falling wave boundary condition near the black hole horizon~\cite{Herzog:2002pc,Crossley:2015tka,Glorioso:2018mmw,deBoer:2018qqm,Chakrabarty:2019aeu,Baggioli:2023tlc}. 
However, our findings arise from imposing the shift symmetry rather than specific traits of black holes.

\vspace{10pt}
\sect{Summary}``Hydrodynamic modes are all alike; every nonhydrodynamic sector is different in its own way.''

Contrary to this common lore of thermalization in quantum field theory, we have shown that diverse nonhydrodynamic excitations—from holographic quasinormal modes to kinetic quasiparticles and their fluctuations—can be unified in a local EFT framework by generalizing the concept of shift symmetry. 
These symmetries underlie the distinct analytic structures of different theories and enable the construction of hybrid models relevant to systems like the asymptotically free QGP liquid. The restrictive shift-symmetry constraint notwithstanding, our framework covers almost all interesting situations discussed in the literature regarding charge current.

Our work is motivated by the apparent importance of nonhydrodynamic modes in characterizing the thermalization in various physical systems~\cite{Romatschke:2017vte,Kurkela:2019set,Brewer:2019oha,Brewer:2022vkq,Ke:2022tqf,Ke:2023bei,An:2023yfq,DeLescluze:2025gaa,DeLescluze:2025jqx}.
The framework presented here provides the basis for the action formulation to describe these modes.

Future directions include extension for the energy-momentum sector~\cite{Liu:2018kfw}, non-Abelian symmetries (e.g. SU(2) in the context of spin hydrodynamics~\cite{Florkowski:2017ruc} and spin kinetic theory~\cite{Mo:2025nes}) and higher spin symmetries~\cite{Giombi:2012ms}. All of these aspects are directly relevant for QCD applications. Our hybrid model can be further developed in more complex scenarios that in turn lead to a richer landscape of nonhydrodynamic excitations~\cite{Hartnoll:2005ju,Grozdanov:2016vgg,Grozdanov:2024fxr}. Finally, the applications of our framework to non-Gaussian fluctuations far from equilibrium are worth exploring~\cite{Jain:2020zhu,An:2020vri,Sogabe:2021svv,Abbasi:2024pwz}.

\vspace{10pt}
\sect{Acknowledgments}
This project has received funding from the European Research Council (ERC) under the European Union’s Horizon 2020 research and innovation programme (grant No.~101089093; project acronym: High-TheQ). Views and opinions expressed are however those of the authors only and do not necessarily reflect those of the European Union or the European Research Council. Neither the European Union nor the granting authority can be held responsible for them. This work was partially supported  by the Priority Research Area Digiworld under the program Excellence Initiative  - Research University at the Jagiellonian University in Krakow.
Y.Y. acknowledges the support from NSFC under Grant No.~12575150 and by CUHK-Shenzhen University Development Fund under Grant No.~UDF01003791 . 
We thank M. Baggioli, J.P. Blaizot, B. Gouteraux, Jin Hu, A. Jain, P. Kovtun, L. Gavassino, Shu Lin, M. Mezei, A. Soloviev, M. Stephanov, D. Teaney, and B. Withers for helpful discussions and/or comments on the draft. Anthropic’s Claude Opus 4.6, and DeepSeek's DeepSeek-V4-Pro 4.0 Pro were helpful in proofreading the
manuscript. We also acknowledge the hospitality of \emph{Galileo Galilei Institute} and \emph{European Center for Theoretical Studies in Nuclear Physics and Related Areas (ECT*)} where some part of the work was completed.

\bibliographystyle{bibstyl}
\bibliography{letter} 

\onecolumngrid
\makeatletter
\begingroup
\vspace{50pt}
  \let\MakeTextUppercase\@firstofone
  \section*{End Matter}
\vspace{10pt}
\endgroup
\makeatother
\twocolumngrid
\renewcommand{\theequation}{\Alph{section}\arabic{equation}}

\vspace{10pt}
\appsect{On the condition $\sA_{1,t}=0$}
Let us now explain the rationale for imposing $\mathcal A_{1,t}=0$. If instead
all $\mathcal A_{n,t}$, $n=1,\dots,K$, are dynamical, the EOMs
for those fields, in the homogeneous and static limit, with
$J_{n,t}\equiv f_n^2\,\Delta\mathcal A_{n,t}$, becomes
\begin{equation}
\label{rho2}
J_{1,t}=0,\quad J_{1,t}=J_{2,t},\quad \ldots,\quad
J_{K-1,t}=J_{K,t}\equiv J_{B,t}.
\end{equation}
The second through $K$th conditions, $J_{n-1,t}=J_{n,t}$, arise from varying
$\mathcal A_{n,t}^a$, $n=2,\dots,K$. They together imply
$J_{1,t}=J_{2,t}=\cdots=J_{K,t}\equiv J_{B,t}$. The first condition, $J_{1,t}=0$,
originates precisely from the EOM for $\mathcal A_{1,t}$, and it
forces the background charge density $J_{B,t}$ to vanish. To allow a finite
background charge for a homogeneous, static state, we must instead hold
$\mathcal A_{1,t}$ fixed.

Eq.~\eqref{rho2} also shows why it is $\mathcal A_{1,t}$, and not another
field, that must be fixed. Since $\mathcal A_{1,t}$ is the end of the
hidden-sector chain, only this end-point field's EOM produces
$J_{1,t}=0$. As for the value zero (the Dirichlet boundary condition for the
chain), we can always reach it by a time-dependent shift, since
$\delta_s\mathcal A_{1,t}=\partial_t\lambda(t,\bm x)$ without changing the physical observables such as $J_{B,t}$.

\vspace{10pt}
\appsect{Discussion on the fluctuation-dissipation part of the action}
We detail here the evaluation of action~\eqref{I-flut} to obtain \eqref{I-FD-full}.
Denoting $\Lambda\equiv \partial_{\theta}\tilde{\chi}$,
the EOM for $\tilde{\chi}$ by varying~\eqref{I-flut} reads
\begin{align}
    i2\pi T'\,\partial_{\theta}\Lambda = \partial_{t}\Lambda\, .
\end{align}
In Fourier space, the solution becomes
\begin{align}
\label{L-sol}
    \Lambda(\theta,\omega) = C(\omega)\, e^{-\frac{\omega}{2\pi T'}\theta}\, .
\end{align}
The integration constant $C(\omega)$ is fixed by matching to the boundary condition
\begin{align}
    \int^{2\pi}_0 d\theta\, \Lambda 
    = \tilde{\chi}(2\pi)-\tilde{\chi}(0)
    = \chi_{+}(\omega)-\chi_{-}(\omega)=\chi_a(\omega)\, .
\end{align}
Evaluating the integral of the solution~\eqref{L-sol}
gives
\begin{align}
\label{C-exp}
    C(\omega) = \frac{\omega\,\chi_a(\omega)}{2\pi T'\big(1-e^{-\omega/T'}\big)}\, .
\end{align}

Upon using the EOM, the on-shell action reduces to the boundary term, which is identified as
\begin{align}
    I_{\rm FD} 
    = i2\pi T'\,\gamma\int_{x,\omega} \Lambda\,\tilde{\chi}\Big|^{\theta=2\pi}_{\theta=0}\, .
\end{align}
Substituting the on-shell solution for $\Lambda$
and using $\chi_{\pm} = \chi_r \pm \chi_a/2$, we obtain after straightforward algebra and relabeling $\omega\to-\omega$ precisely the expression~\eqref{I-FD-full} with the identification $T=T'$.
Note in the low-frequency limit, it reduces to the familiar form
\begin{align}
    I_{\rm FD}\approx
    i\int_{x,\omega}\gamma\,
    \chi_a(-\omega)\Big[\omega\,\chi_r(\omega) + T\,\chi_a(\omega)\Big]\, .
\end{align}

\vspace{10pt}
\appsect{Continuum Limit and Holographic Matching of $I_{\rm B}$}\label{app:continuum}
We derive the continuum limit of $I_{\rm B}=I_{B,a}+I_{B,aa}$ 
in the symmetry-broken case, and demonstrate that it can be matched to the 
holographic SK action under the prescription specified below. 
We now suppress the subscript $B$ throughout.

We decompose the full action into a nondissipation (ND) 
and a fluctuation-dissipation (FD) part.  
The nondissipation part is obtained from Eq.~\eqref{I-HLS} by 
dropping terms proportional to $\sigma_n$, and factorizes as
\begin{align}
    I_{\rm ND}=I_{0}[\mathcal{A}^{+}]-I_0[\mathcal{A}^{-}]\,,
\end{align}
where
\begin{align}
\label{I-HLS-0}
I_{0}[\mathcal{A}]=\frac{1}{2}\int_x \sum_{n=1}^{K}  
     (f^2_n\,\Delta \mathcal{A}^2_{n,t}
     - g^2_{n}\, \Delta\bm{\mathcal{A}}^2_{n} 
    +\epsilon_n \bm{E}^2_n
    -\kappa_n\bm{B}^2_{n}).
\end{align}

To take $K\to\infty$ limit, we introduce the infinitesimal layer/site spacing 
$\Delta\rho$ and define rescaled couplings that remain finite 
as $\Delta\rho\to 0$,
\begin{align}\label{eq:rescale-fgek}
    \bigl(f(\rho_n)^2,\,g(\rho_n)^2\bigr)
        &=\bigl(f^2_n,\,g^2_n\bigr)\Delta\rho\,,\no\\
    \bigl(\epsilon(\rho_n),\,\kappa(\rho_n)\bigr)
        &=\bigl(\epsilon_n,\,\kappa_n\bigr)/\Delta\rho\,.
\end{align}
No rescaling is needed for the gauge field:
$\mathcal{A}_{n,\mu}=A_\mu(\rho_n)$,\;
$\bm{E}_n=\bm{E}(\rho_n)$,\;
$\bm{B}_n=\bm{B}(\rho_n)$.
Through $\rho_n=n\Delta\rho+\rho_0$, with $\rho_0$ being the coordinate of the boundary where $\sA_1$ lives, all $n$-dependent couplings and gauge fields become 
smooth functions of the fifth-dimensional coordinate $\rho$.
At the same time the value of $\sA$ at $\rho_{K}$ precisely gives that of the external gauge field.

Introducing a 5D gauge field $A_M=(A_\mu,A_\rho)$ via 
$\Delta\varphi_n=-\Delta\rho\,A_\rho(\rho_n)$, the field strength 
$F_{MN}\equiv\partial_M A_N-\partial_N A_M$ leads to the relation $\Delta \mathcal{A}_{n,\mu}=F_{\rho\mu}(\rho_n)\,\Delta\rho $.
Now replacing the sum with an integral yields
\begin{align}
\label{eq:I0-continuum}
    I_{0}[\mathcal{A}] = \frac{1}{2}\int_{x,\rho}
    \Bigl(
    f^2 F^2_{\rho t} 
    - g^2 F^2_{\rho i} 
    + \epsilon\, F^2_{it} 
    - \tfrac{1}{2}\kappa\, F^2_{ij}
    \Bigr)\,.
\end{align}
Eq.~\eqref{eq:I0-continuum} is identified with the 
Yang-Mills action in $4+1$D bulk spacetime,
\begin{equation}
\label{eq:I0-YM}
    I_{0} = -\frac{1}{4}\int_{x,\rho} \sqrt{-\mathsf{g}}\,
     \mathsf{g}^{MM'}\mathsf{g}^{NN'}F_{MN}F_{M'N'}\,.
\end{equation}
Here, the diagonal metric 
$ds^2=-g_{tt}dt^2+g_{ii}d\bm{x}^2+g_{\rho\rho}d\rho^2$ is determined by
\begin{gather}
    \sqrt{-\mathsf{g}}\,\mathsf{g}^{\rho\rho}\mathsf{g}^{tt}
        =f^2(\rho)\,,\quad
    \sqrt{-\mathsf{g}}\,\mathsf{g}^{\rho\rho}\mathsf{g}^{ii}
        =g^2(\rho)\,, \no\\
    \sqrt{-\mathsf{g}}\,\mathsf{g}^{ii}\mathsf{g}^{tt}
        =\epsilon(\rho)\,,\quad
    \sqrt{-\mathsf{g}}\,\mathsf{g}^{ii}\mathsf{g}^{jj}
        =\kappa(\rho)\,.
    \label{eq:metric-match}
\end{gather}
Hence, the discrete hidden-layer couplings in continuum limit directly encode an 
emergent gravitational geometry.

Now we turn to the FD part of the action, which encodes both dissipation 
and statistical fluctuations.
For definiteness, we shall assume that the only nonzero $\sigma_1=\sigma$.
Applying the general expression~\eqref{I-flut} to the last term in Eq.~\eqref{I-HLS}, we obtain
\begin{align}
\label{eq:IFD}
    I_{\rm FD}=\int_{x,\theta}\sigma\,\left[
    2\pi Ti(\partial_{\theta}{\bm{\mathcal{\tilde{A}}}})^2
    -(\partial_{\theta}
    {\bm{\mathcal{\tilde{A}}}})\cdot\partial_{t}{\bm{\mathcal{\tilde{A}}}}\right]\,.
\end{align}

Finally, we demonstrate that the action \eqref{eq:I0-continuum} together with \eqref{eq:IFD} can be matched to the holographic description of gauge plasma.
According to the holographic prescription for the thermal system, the metric has to have a horizon at $\rho_0$.
So it is convenient to consider the ingoing Eddington-Finkelstein (EF) coordinate $v$ where
\begin{align}
    dv= dt +\sqrt{\frac{{\mathsf g}_{\rho\rho}}{{\mathsf g}_{tt}}}\,d\rho\, ,
\end{align}
and the metric takes the form
\begin{align}
    ds^{2} = -{\mathsf g}_{tt}\,dv^{2} + 2\sqrt{{\mathsf g}_{tt}{\mathsf g}_{\rho\rho}}\;dv\,d\rho + {\mathsf g}_{ii}\,d\bm{x}^{2}.
\end{align}

For illustrative purpose, we focus on the part of the action that depends on ${\bm \sA}$ and will consider the situation that ${\bm \sA}$ has no spatial dependence.
The relevant action in EF coordinates then takes the form
\begin{equation}
\label{eq:I0-YM-EF}
I_{0}[\mathcal{A}] = \frac12\int_{\vx,v,\rho}\;\sqrt{-\mathsf g}
\Bigl[ - \frac{2\,\mathsf g^{ii}}{\sqrt{g_{tt} g_{\rho\rho}}}\,F_{vi}F_{\rho i}
      - \frac{\mathsf g^{ii}}{\mathsf g_{\rho\rho}}\,(F_{\rho i})^{2} \Bigr]\,.
\end{equation}

Near the horizon, the metric functions behave universally as
\begin{align}
    \mathsf g_{tt} \approx c_t (\rho-\rho_0), \qquad 
    \mathsf g_{\rho\rho} \approx \frac{c_\rho}{\rho-\rho_0} \, ,
\end{align}
where $c_t,c_\rho$ are constants.
Particularly, 
the Hawking temperature is given by $4\pi T=\sqrt{c_t/ c_\rho}$.
The near-horizon action now reads
\begin{equation}\label{eq:action-T}
I_{0}[\boldsymbol{\mathcal{A}}] 
= -\sqrt{\mathsf g^{ii}(\rho_0)}\int_{v,\rho}\,
\,\partial_{v}\boldsymbol{\mathcal{A}}\cdot\partial_{\rho}\boldsymbol{\mathcal{A}}
      + 2\pi T\,(\rho-\rho_{0})\,(\partial_{\rho}\boldsymbol{\mathcal{A}})^{2}.
\end{equation}

To make a connection to the action~\eqref{eq:IFD}, we consider action~\eqref{eq:action-T} for the field lives on a small circle in $\rho$ space and is parametrized by $\theta$:
\(\rho-\rho_{0}= r e^{i\theta}\) for an infinitesimal $r$. 
The radial derivative and measure become
$
\partial_{\rho} =-\frac{i}{r}e^{-i\theta}\partial_{\theta},
d\rho = i r e^{i\theta}\,d\theta$,
and the action~\eqref{eq:action-T} indeed reduces to the action~\eqref{eq:IFD} with \(\sigma = \sqrt{\mathsf g^{ii}(\rho_0)}\) if we further identify $t$ in \eqref{eq:IFD} with time in EF-coordinate $v$.
This identification, though, looks quite natural. 

\begin{figure}[t]
 \centering
  \includegraphics[width=0.48\textwidth]{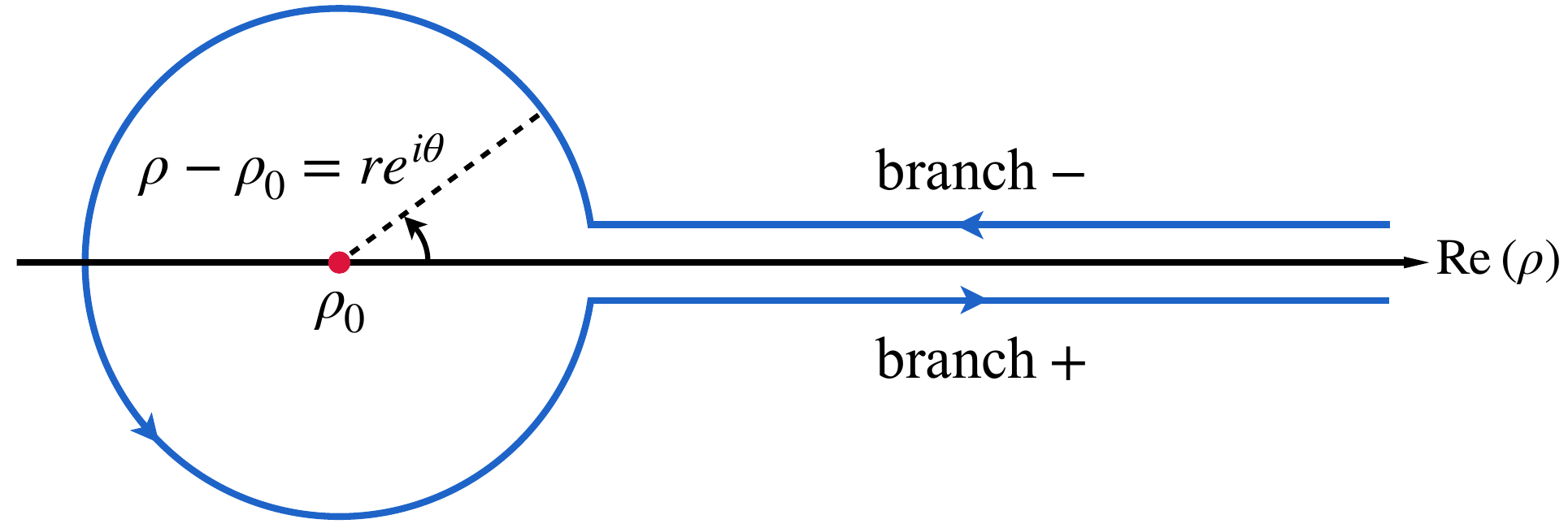}
  \caption{Illustration of the SK contour.}
  \label{fig:contour}
\end{figure}

To complete the matching, we still need to impose the boundary in accordance with Eq.~\eqref{chi-bound},
\begin{equation}
\label{A-match}
    \tilde{\mathcal{A}}(\theta=0)=\mathcal{A}^{-}\,,
    \qquad
    \tilde{\mathcal{A}}(\theta=2\pi)=\mathcal{A}^{+}\, .
\end{equation}
This boundary condition has the following interpretation for the full action in the continuum limit,
\begin{align}
    I_{B}=I_0[\sA^{+}]-I_0[\sA^{-}]+I_{\rm FD}[\tilde{\sA}]\,,
\end{align}
for the complexified contour as illustrated in Fig.~\ref{fig:contour}. The contour consists of two parallel paths connecting $\rho_{K}$ and $\rho_0$, and the fields on the first and second legs are represented by $\sA^{-}$ and $\sA^{+}$, respectively. Meanwhile, the field on the circle that connects two legs is denoted by $\tilde{\sA}$, as we have just demonstrated. It is now straightforward to verify that this action is equivalent to the holographic SK action defined on the contour in Ref.~\cite{Glorioso:2018mmw}.

\end{document}